\documentclass[graybox]{svmult}

\usepackage{helvet}         
\usepackage{courier}        
\usepackage{type1cm}        
\usepackage{makeidx}         
\usepackage{graphicx}        
\usepackage{multicol}        
\usepackage[bottom]{footmisc}

\makeindex             

\begin{document}

\title*{Galactic Plane H$\alpha$ Surveys: IPHAS \& VPHAS+}
\author{Nicholas J. Wright}
\institute{Nicholas J. Wright, on behalf of the IPHAS and VPHAS+ consortia, see the survey websites www.iphas.org and www.vphasplus.org for a full list of collaborators. \at Centre for Astrophysics Research, University of Hertfordshire, \email{nick.nwright@gmail.com}}

\maketitle

\abstract{
The optical Galactic Plane H$\alpha$ surveys IPHAS and VPHAS+ are dramatically improving our understanding of Galactic stellar populations and stellar evolution by providing large samples of stars in short lived, but important, evolutionary phases, and high quality homogeneous photometry and images over the entire Galactic Plane. Here I summarise some of the contributions these surveys have already made to our understanding of a number of key areas of stellar and Galactic astronomy.
}

\section{Introduction}

H$\alpha$ emission is an important tracer of many critical phases of stellar evolution, including pre- and post-main-sequence (MS) stars, interacting binaries and massive stars, as well as tracing diffuse ionised nebulae. Despite their importance these evolutionary phases are generally short-lived, and therefore stars in these phases are rare and difficult to find. The Galactic Plane surveys IPHAS (INT Photometric H$\alpha$ Survey, \cite{drew05}) and VPHAS+ (VST Photometric H$\alpha$ Survey \cite{drew14}) represent a concerted effort to overcome this deficiency, to improve our understanding of these short-lived evolutionary phases and to facilitate large-scale stellar population and Galactic structure studies.

The two surveys are imaging the entire Northern (IPHAS) and Southern (VPHAS+) Galactic Plane and bulge using the Isaac Newton Telescope (INT) and VLT Survey Telescope (VST). Together with their blue-filter partner survey UVEX (UV Excess Survey \cite{groo09}) on the INT they are covering an area of $\sim$3800 sq. deg., across the entire Galactic Plane ($0^\circ \leq l < 360^\circ$, $|b| < 5^\circ$) and bulge ($10^\circ > l > 350^\circ$, $|b| < 10^\circ$), using the filters $u$, $g$, $r$, $i$, and H$\alpha$. The H$\alpha$ filters used by IPHAS ($\lambda_c = 6568$~\AA) and VPHAS+ ($\lambda_c = 6589$~\AA) are broad enough (FWHM of 95~\AA\ and 100~\AA\ respectively) to capture most Doppler shifts due to Galactic motion.

The observations are taken in blue ($u/g/r$) and red ($r/i/$H$\alpha$) blocks, providing co-eval multi-band photometry necessary for compiling the many colour-colour diagrams facilitated by the surveys that can efficiently be used to identify and characterise rare types of object. The observations have been performed under good conditions (VPHAS+ median seeing $=$ 0.8$^{\prime\prime}$) and reach a 5$\sigma$ depth of $g, r \sim 22$~mag, and saturation at $\sim$11--13~mag. The IPHAS second data release \cite{bare14} can be obtained from the survey website (www.iphas.org/dr2), while VPHAS+ data is available from the ESO science archive (http://www.eso.org/sci/observing/phase3/data\_releases.html).

\section{Star formation and young stars}

\begin{figure}[t]
\sidecaption[t]
\includegraphics[width=7.5cm]{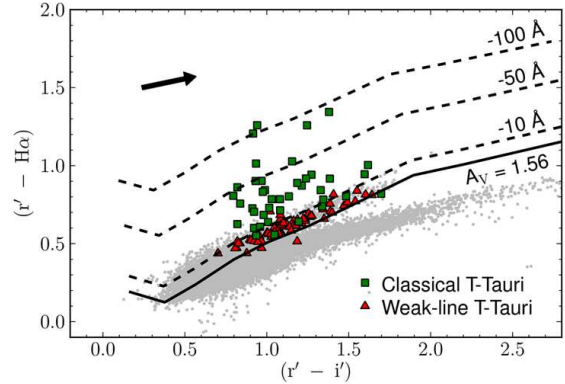}
\caption{The IPHAS ($r^\prime - \mathrm{H}\alpha$) vs ($r^\prime - i^\prime$) colour-colour diagram showing the identification of H$\alpha$-emitting classical T-Tauri stars in IC~1396 relative to known weak-line T-Tauri stars. Also shown are a reddened main-sequence (solid line) and the positions of stars at increasing levels of H$\alpha$ emission. Grey dots shown field stars in the region. Figure from \cite{bare11}.}
\label{HAselection}
\end{figure}

H$\alpha$ emission is a common feature of pre-MS stars due to accretion from a circumstellar disc onto the star itself, and therefore can be used as a diagnostic of youth \cite{vink08,with08}. Furthermore the $r - $H$\alpha$ colour also provides a measure of the equivalent width of the H$\alpha$ emission line (see Fig~\ref{HAselection}), which can then provide an estimate of the accretion rate. Barentsen et al. \cite{bare11} used IPHAS photometry for stars towards the H~{\sc ii} region IC~1396, identifying 158 pre-MS candidates with masses between 0.2 and 2.0~M$_\odot$. They measured accretion rates for these stars and found a power-law dependency between the stellar mass and the accretion rate with a slope of $\alpha = 1.1 \pm 0.2$. They also found evidence for an age gradient, manifested through accretion rates and infrared excesses, with younger stars lying further from the central ionising O-type star of the region, which provides evidence that the formation of these stars was sequentially triggered by the O star.

The H$\alpha$ imagery from these surveys also facilitates large-scale studies of the diffuse emission across H~{\sc ii} regions, and has already led to the discovery of a unique class of young stellar object experiencing feedback from nearby OB stars \cite{wrig12}.

\section{Massive stars}

\begin{figure}[t]
\sidecaption[t]
\includegraphics[width=7.0cm]{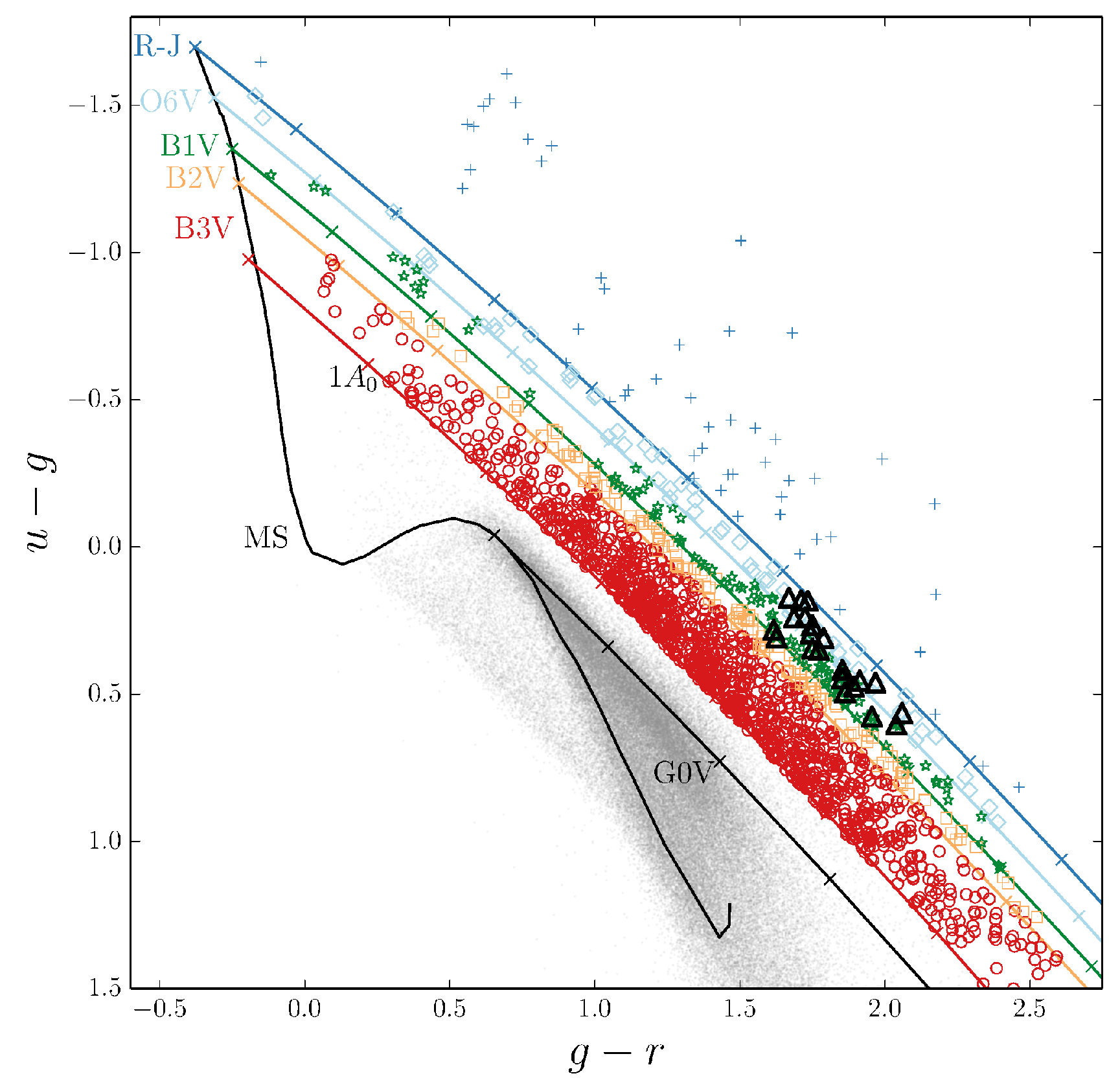}
\caption{The VPHAS+ ($u-g$) vs ($g-r$) colour-colour diagram showing the identification of O and B stars (coloured circles, as per the spectral types listed) and known OB stars in Westerlund~2 (black triangles). Also shown are an unreddened main sequence (MS, black line) and $R_V = 3.8$ reddening lines (coloured lines) extending from various labelled spectral types and from the position of a Rayleigh-Jeans spectrum. Figure adapted from \cite{mohr15}.}
\label{OBselection}
\end{figure}

Massive O and early B stars can be efficiently identified and characterised using the blue filters employed by the UVEX and VPHAS+ surveys (see Fig~\ref{OBselection}). Mohr-Smith et al. \cite{mohr15} have used this method to identify 356 new O and early B star candidates across a 2 sq. degree region of the Carina spiral arm of the Milky Way at distances of 3--6~kpc. The photometry can be used to fit spectral types and reddening parameters, which were verified using known OB stars in the field. This allowed the authors to map out the distribution of massive stars across the region and study the variation of the reddening law, which requires non-standard values of $3.5 < R_V < 4$.

The H$\alpha$ imagery from these surveys can also be used to study the environments of massive stars to probe their mass-loss history. Wright et al. \cite{wrig14} presented H$\alpha$ images of an ionised nebula surrounding the red supergiant (RSG) W26 in the massive star clusters Westerlund~1 consisting of a circumstellar shell or ring $\sim$0.1~pc in diameter and a triangular nebula $\sim$0.2~pc from the star. The authors hypothesise that the circumstellar nebula likely formed from a previous episode of mass-loss from the RSG, while the triangular nebula may be a flow of material channelled off the circumstellar nebula by the cumulative cluster wind and radiation field.

\section{Evolved stars and stellar remnants}

The filters employed by these surveys are very effective for identifying evolved stars and stellar remnants such as asymptotic giant branch stars \cite{wrig08}, white dwarfs \cite{verb13}, supernova remnants \cite{sabi13}, and planetary nebulae (PNe) \cite{sabi10}. In the latter category the H$\alpha$ filter employed by these surveys has been valuable for discovering many new compact \cite{viir09} and extended \cite{sabi14} PNe, as well as improving our understanding of those already known \cite{ware06}. The blue filters employed by these surveys also allows the central stars of many existing PNe to be identified for the first time \cite{drew14}. Finally the legacy value of these deep and detailed surveys should not be underestimated, particularly as we enter an era giving more emphasis to time-domain studies, such as for identifying and studying the progenitors of novae \cite{wess08}.

\section{Galactic structure}

The combination of precise and high spatial resolution photometry, the unique ability of $r-$H$\alpha$ to discriminate intrinsic stellar colour, and the wide area coverage facilitates detailed studies of the spatial distribution of stars, gas and dust across our Galaxy. 
Farnhill et al. ({\it in prep}) produced calibrated stellar density maps from IPHAS data to test Galactic population synthesis models and 3D extinction maps. The authors found a number of discrepancies between the models and observations, with the main problem being an under-prediction of extinction at low Galactic longitudes ($l \simeq 30^\circ$). 
Sale et al. \cite{sale14} used IPHAS photometry and a hierarchical Bayesian model to compile a 3D map of extinction across the northern Galactic plane at an unprecedented spatial ($\sim$10~arcmin) and distance (100~pc) resolution that will be vital for studies of Galactic stellar populations. 

\section{Summary}

The H$\alpha$ Galactic Plane surveys IPHAS and VPHAS+ have brought many areas of stellar and Galactic astronomy into the modern era by providing high quality and homogeneous photometry for hundreds of millions of stars across our Galaxy. The science discussed here highlights the contribution these surveys have already made to many key areas of astrophysics. Once completed these surveys will provide a valuable resource for many future studies, particularly when combined with astrometry from ESA's Gaia satellite.

\begin{acknowledgement}
NJW acknowledges a Royal Astronomical Society Research Fellowship and is grateful to all those within IPHAS and VPHAS+ who have made these surveys possible and contributed to them. Thanks to M. Mohr-Smith for Figure~2.
\end{acknowledgement}


\begin{thebibliography}{99.}

\bibitem{bare11} Barentsen, G., Vink, J. S., Drew, J. E., et al. T Tauri candidates and accretion rates using IPHAS: method and application to IC~1396. MNRAS \textbf{415}, 103 (2011)

\bibitem{bare14} Barentsen, G., Farnhill, H. J., Drew, J. E., et al. The second data release of the INT Photometric H$\alpha$ Survey of the Northern Galactic Plane (IPHAS~DR2). MNRAS \textbf{444}, 3230 (2014)

\bibitem{drew05} Drew, J.~E., Greimel, R., Irwin, M. J., et al. The INT Photometric H$\alpha$ Survey of the Northern Galactic Plane (IPHAS). MNRAS \textbf{362}, 753 (2005)

\bibitem{drew14} Drew, J.~E., Gonzalez-Solares, E., Greimel, R.,et al. The VST Photometric H$\alpha$ Survey of the Southern Galactic Plane and Bulge (VPHAS+). MNRAS \textbf{440}, 2036 (2014)

\bibitem{groo09} Groot, P.~J., Verbeek, K., Greimel, R., et al. The UV-Excess survey of the Northern Galactic Plane (UVEX). MNRAS \textbf{399}, 323 (2009)

\bibitem{mohr15} Mohr-Smith, M., Drew, J. E., Barentsen, G.,et al. New OB star candidates in the Carina Arm around Westerlund 2 from VPHAS+. MNRAS, {\it in press}, 2015, arXiv 1504.04342.

\bibitem{sabi10} Sabin, L., Zijlstra, A. A., Wareing, C., et al. New Candidate Planetary Nebulae in the IPHAS Survey: the Case of Planetary Nebulae with ISM interaction. PASA \textbf{27}, 166 (2010).

\bibitem{sabi13} Sabin, L., Parker, Q. A., Contreras, M. E., et al. New Galactic supernova remnants discovered with IPHAS. MNRAS \textbf{431}, 279 (2013)

\bibitem{sabi14} Sabin, L., Parker, Q. A., Corradi, R. L. M., et al. First release of the IPHAS catalogue of new extended planetary nebulae. MNRAS \textbf{443}, 3388 (2014)

\bibitem{sale14} Sale, S. E., Drew, J. E., Barentsen, G., et al. A 3D extinction map of the northern Galactic plane based on IPHAS photometry. MNRAS \textbf{443}, 2907 (2014)

\bibitem{verb13} Verbeek, K., Groot, P. J., Nelemans, G., et al. A determination of the space density and birth rate of hydrogen-line (DA) white dwarfs in the Galactic plane, based on the UVEX survey. MNRAS \textbf{434}, 2727 (2013)

\bibitem{viir09} Viironen, K., Greimel, R., Corradi, R. L. M., et al. Candidate planetary nebulae in the IPHAS photometric catalogue. A\&A \textbf{504}, 291 (2009)

\bibitem{vink08} Vink, J.~S., Drew, J. E., Steeghs, D., et al. IPHAS discoveries of young stars towards Cyg~OB2 and its southern periphery. MNRAS \textbf{387}, 308 (2008)

\bibitem{ware06} Wareing, C.~J., O'Brien, T. J., Zijlstra, A. A., et al. The shaping of planetary nebula Sh2-188 through interaction with the interstellar medium. MNRAS \textbf{366}, 387 (2006)

\bibitem{wess08} Wesson, R., Barlow, M. J., Corradi, R. L., M., et al. A planetary nebula around Nova V458 Vulpeculae undergoing flash ionization. ApJL \textbf{688}, 21 (2008)

\bibitem{with08} Witham, A.~R., Knigge, C., Drew, J. E., et al. The IPHAS catalogue of H$\alpha$ emission-line sources in the northern Galactic plane. MNRAS \textbf{384}, 1277 (2008)

\bibitem{wrig08} Wright, N.~J., Greimel, R., Barlow, M. J., et al. Extremely red stellar objects revealed by IPHAS. MNRAS \textbf{390}, 929 (2008)

\bibitem{wrig12} Wright, N.~J., Drake, J. J., Drew, J. E., et al. Photoevaporating proplyd-like objects in Cygnus OB2. ApJL \textbf{746}, 21 (2012)

\bibitem{wrig14} Wright, N.~J., Wesson, R., Drew, J. E., et al. The ionised nebula surrounding the red supergiant W26 in Westerlund 1. MNRAS \textbf{437}, 1 (2014)

\end{thebibliography}
\end{document}